\documentclass[manuscript,screen]{acmart}
\AtBeginDocument{%
  \providecommand\BibTeX{{%
    \normalfont B\kern-0.5em{\scshape i\kern-0.25em b}\kern-0.8em\TeX}}}

\setcopyright{acmcopyright}
\copyrightyear{2023}
\acmYear{2023}
\acmDOI{XXXXXXX.XXXXXXX}

\usepackage{comment}

\begin{document}

\title{Gaze-based Attention Recognition for Human-Robot Collaboration}

\author{Pooja Prajod}
\email{pooja.prajod@uni-a.de}
\orcid{0000-0002-3168-3508}
\affiliation{%
  \institution{Universit\"at Augsburg}
  \streetaddress{Universit\"atsstr. 6a}
  \city{Augsburg}
  \state{Bavaria}
  \country{Germany}
  \postcode{86159}
}

\author{Matteo Lavit Nicora}
\affiliation{%
  \institution{National Research Council of Italy}
  \city{Lecco}
  \country{Italy}}
\email{matteo.lavit@stiima.cnr.it}

\author{Matteo Malosio}
\affiliation{%
  \institution{National Research Council of Italy}
  \city{Lecco}
  \country{Italy}}
\email{matteo.malosio@stiima.cnr.it}

\author{Elisabeth Andr\'e}
\affiliation{%
  \institution{Universit\"at Augsburg}
  \city{Augsburg}
  \country{Germany}
}
\email{elisabeth.andre@uni-a.de}

\renewcommand{\shortauthors}{Prajod, et al.}

\begin{abstract}
Attention (and distraction) recognition is a key factor in improving human-robot collaboration.
We present an assembly scenario where a human operator and a cobot collaborate equally to piece together a gearbox.
The setup provides multiple opportunities for the cobot to adapt its behavior depending on the operator's attention, which can improve the collaboration experience and reduce psychological strain.
As a first step, we recognize the areas in the workspace that the human operator is paying attention to, and consequently, detect when the operator is distracted.
We propose a novel deep-learning approach to develop an attention recognition model.
First, we train a convolutional neural network to estimate the gaze direction using a publicly available image dataset.
Then, we use transfer learning with a small dataset to map the gaze direction onto pre-defined areas of interest.
Models trained using this approach performed very well in leave-one-subject-out evaluation on the small dataset.
We performed an additional validation of our models using the video snippets collected from participants working as an operator in the presented assembly scenario.
Although the recall for the Distracted class was lower in this case, the models performed well in recognizing the areas the operator paid attention to.
To the best of our knowledge, this is the first work that validated an attention recognition model using data from a setting that mimics industrial human-robot collaboration.
Our findings highlight the need for validation of attention recognition solutions in such full-fledged, non-guided scenarios.
\end{abstract}




\maketitle

\section{Introduction}

In Industry 4.0 scenarios, a human operator's gaze reveals information that a collaborating robot (cobot) may use to adjust its behavior. For example, if the operator is distracted from the assembly, the cobot partner could slow down or place the assemblies in a dedicated location to avoid disruption of the assembly line.
Such human-centered adaptations can reduce negative experiences and psychological strain on the operator~\cite{bib:mindbot}.

Gaze is an important social cue that has been demonstrated to facilitate collaboration~\cite{gazegrounding,huang2016anticipatory, palinko2016robot}.
It can be leveraged for resolving ambiguities, joint attention, etc.~\cite{gazegrounding}.
In this work, we focus on gaze direction, which communicates the current area of interest to the collaborating partners~\cite{dualreality}.
Previous studies~\cite{dualreality, saran2018human, shi2021gazeemd,huang2016anticipatory} primarily focused on identifying an object the user is looking at. 
Inspired by the field of driver attention detection~\cite{distracteddriver, vora2017generalizing, tayibnapis2018driver}, we use gaze direction for recognizing attention and distraction of the operators.
This is a first step towards adapting cobot behavior (e.g. speed, specific action sequence) depending on the operator's attention, which is a largely unexplored topic.

To recognize the attention of the operator, we first estimate their gaze direction using a frontal camera and then map the gaze to a pre-defined area of interest.
We propose a novel deep-learning approach to realize this.
First, we train a gaze estimation model using a large publicly available dataset.
Then, we use transfer learning techniques to map the gaze directions to areas of interest.

Previous studies~\cite{saran2018human, tayibnapis2018driver, huang2016anticipatory} have proposed camera-based solutions to identify the area/object that has the user's attention.
A gap in the validation of these systems stems from the guided gaze behaviors in the setup.
The setup typically involves a stationary participant (sitting or standing) who is asked to gaze at a labeled area.
This heavily reduces variations in viewing angle, head poses, etc.
Even in the driver attention use case, where the user is expected to be seated, Ahlstrom et al.~\cite{distracteddriver} point out that achieving ``true distraction'' in an artificial setting is difficult.
In this work, we evaluate our attention recognition model using videos from a human-robot collaboration task that resembles an industrial assembly.
Our observations highlight the importance of testing such models in a full-fledged setup than a well-controlled setting.
To the best of our knowledge, this is the first work to evaluate attention recognition in a setup that imitates an industrial human-robot collaboration task.

\section{Related Work}
\label{sec:related}

Gaze-based attention recognition has been studied thoroughly for various application domains such as driver assistance, and education. 
However, in Human-Robot collaborations (HRC), the gaze is primarily used to detect the object the human partner is looking at.
There is a lack of literature on using gaze to recognize the attention of the human partner or detect the activity they are carrying out.
Consequently, the topic of how the robot should respond is also largely unexplored.

In this section, we briefly discuss a few of the works which guided our research.
We focus mainly on works with HRC use cases, although their goal is identifying the gaze object.
We also discuss a few works from the driver attention detection domain since image/video-based attention recognition is prevalent in this field.

Many works have demonstrated the advantages of using gaze information in HRC scenarios.
Mehlmann et al.~\cite{gazegrounding} demonstrated that gaze can be used during collaboration to resolve ambiguities and to establish joint attention.
They designed a collaborative puzzle game involving a social robot (Nao) and a human partner.
When the robot followed the user's gaze, the interaction was more efficient and natural.
Huang and Mutlu~\cite{huang2016anticipatory} demonstrated that human-robot collaboration can be improved when the robot can perceive the gaze object of the user. 
They designed a scenario where the user had to choose four items from a menu, which the robotic arm then proceeded to pick.
Monitoring the user's gaze enabled the robot to anticipate their choices, and thus pick the object faster.
Shi et al.~\cite{shi2021gazeemd} conducted a similar study, involving the robotic arm picking up an object chosen by the user.
While the previous study involved verbalizing the user's choice, Shi et al. used gaze direction and fixation to determine the user's choice.

Moniri et al.~\cite{dualreality} developed a prototype involving a robot, a local user, and a virtual reality user.
The information about the object that a user is gazing at is provided in both virtual reality and the real setting.
The prototype is an interesting Industry 4.0 scenario but, there was no user study.

The above works use eye-tracking devices to recognize the gaze direction of the user. 
Although such devices can accurately measure the gaze direction, they are often intrusive and are not ideal for real-time applications~\cite{saran2018human, de2022deep}.
This has motivated increasing research interest in gaze estimation using camera images/videos.
There are mainly two ways to recognize gaze direction using a camera image - using eye information and using head pose
For example, Lemaignan et al.~\cite{lemaignan2016real} used head pose in a robot-child collaborative learning scenario to identify the objects that the child paid attention to.
Palinko et al.~\cite{palinko2016robot} compared the two methods of predicting gaze direction based on camera input.
They designed an experiment where the participants had to assemble a tower using four blocks.
The gaze behavior of the participant was used to select the blocks.
Eye-based gaze estimation led to more efficient collaboration.

Christiernin~\cite{collablevels} defines three levels of collaboration - Idle Robot (Level 1), Human as Guide (Level 2), and Cooperation/Full Interaction (Level 3).
All the aforementioned works fall into Level 1 or 2 which makes the collaboration heavily imbalanced i.e., most of the task was carried out by only one of the partners.
This limits certain aspects of collaboration (e.g. waiting for the other partner to finish) and the associated gaze behavior.
In this work, we use an HRC setup that achieves Level 3 collaboration.

Driver distraction detection has gained a lot of traction over the years. 
The driver's attention or distraction is determined based on gaze zones in the car. 
Studies usually use datasets recorded using a frontal camera and involve participants looking at pre-defined gaze zones.
Although the setting and challenges of driver distraction are different, the techniques used to solve this problem could be leveraged for attention recognition in HRC.

Vora et al.~\cite{vora2017generalizing} collected an image dataset with gaze zone labels for detecting driver distraction.
Using this data, they trained two convolutional neural networks.
They achieved good accuracy but, like typical deep learning networks, they required a large amount of data. 
To circumvent this, Tayibnapis et al.~\cite{tayibnapis2018driver} adopted a transfer learning approach.
They connected a pre-trained neural network to a support vector machine for classifying face images to gaze zones. 
The neural network was trained on a dataset of users' gaze toward different types of Apple phones and tablets.
They trained the SVM part of their system using a small dataset that they collected.
Although they achieved very high results on k-fold cross-validation, their model struggled with data from unseen users.

\section{Assembly Task}
\label{sec:task}
\begin{figure}[thpb]
    \includegraphics[width=0.80\textwidth]{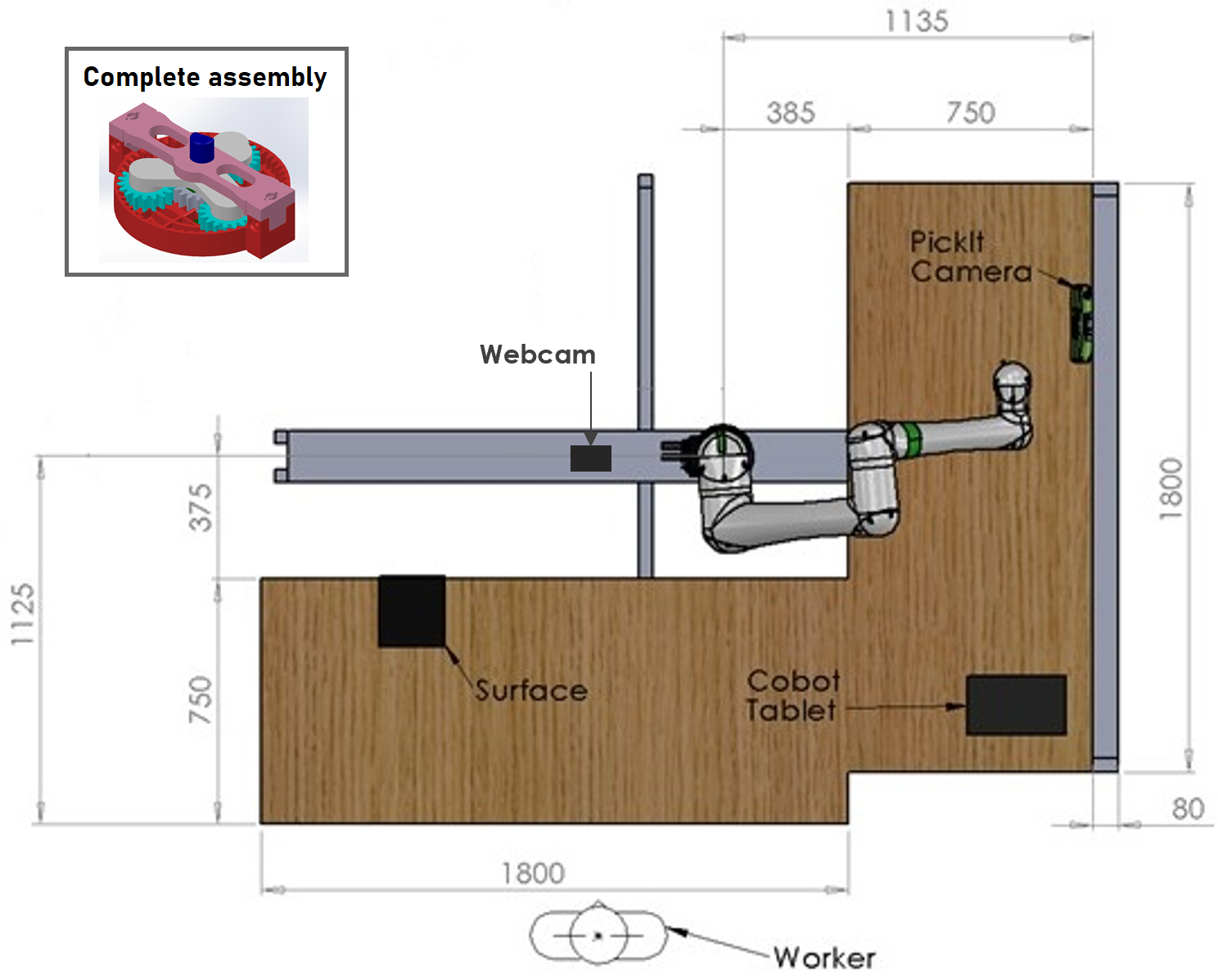}
    \caption{A schematic overview of the experimental workspace with a representation of the complete sub-assembly at the top left corner.}
    \label{fig:cell}
\end{figure}

In order to train and test the gaze-based attention recognition model proposed in this study, the collaborative assembly scenario prepared for the study described in~\cite{bib:mindbot} is exploited. 
The product to be assembled is the 3D-printed planetary gearbox represented in the top left corner of Figure~\ref{fig:cell}. 
Half of the components are put together by the cobot, while the remaining parts are assembled by the human operator. 
The two sub-assemblies are then joint collaboratively in order to obtain the finished product. 
As visualized in Figure~\ref{fig:cell}, a Fanuc CRX10iA/L collaborative robot is mounted on a structure specifically built to guarantee a fixed relative position with respect to two tables arranged in an L-shaped formation. 
The table on the right is equipped with all the components required for the sub-assembly assigned to the cobot, together with a Pickit3D camera, used for the detection of parts. 
The table on the left is where the operator performs most of the activities and also where the collaborative session of the task takes place.
The whole system is driven by a control architecture integrating ROS~\cite{bib:ros}, for controlling both the detection camera and the cobot. 

A simple pilot of the setup and of the assembly task is used to get a rough estimation of the duration of a single production cycle and of the time synchronization with the cobot. 
Generally, the complete cycle takes around 60-70 seconds with the operator finishing first and waiting for the cobot for around 10-15 seconds before tackling the collaborative assembly of the two sub-assemblies.

The assembly is shared equally between the cobot and human operator, making this setup a level 3 (highest level) collaboration according to Christiernin's categorization~\cite{collablevels}.
With the advent of Industry 4.0, this level of collaboration is becoming more common in the manufacturing process.
Thus, this setup allows us to explore how the collaborative experience of the operator can be improved.
Similar to an industrial assembly scenario, the interaction is bound to the defined task and production protocol~\cite{bib:mindbot}.
In other words, we can adapt how the cobot does the task (e.g. adapting speed) but the task load remains more or less the same.

\section{Gaze-Based Attention Recognition}
While piloting our setup, we observed that there are two key areas that the operator pays attention to for an extended period of time - cobot and work table.
We also envision the operator getting distracted when working for long hours.
So, we define three classes of attention based on the gaze of the operator - attention on the cobot, attention on the table, and distracted (looking in some other direction).

Recognition of these areas can be valuable in reducing the psychological strain on the operator and in turn improve the collaboration experience.
It is a reasonable heuristic that if the operator's attention is on the table, they are working on the sub-assembly.
Similarly, if they are looking at the robot, they are plausibly waiting for the robot to bring its sub-assembly.
If the cobot assembles faster and visibly waits for the operator, the operator might feel pressured to speed up in order to synchronize with the cobot.
So, if the operator is still assembling (attention to the table), the robot should wait inconspicuously or proceed to assemble the next part till the operator is ready.
On the other hand, if the operator is faster and is waiting for the cobot (look at the cobot), then the cobot should increase its pace to avoid boredom.
Our ultimate goal is to enable the cobot to adapt its behavior in response to the operator's social and affective cues (e.g. gaze).
As a first step towards this goal, we train a gaze-based attention recognition model to detect the area the user is currently focusing on.

We consider a transfer learning approach to train our attention recognition model.
Transfer learning involves using the parameters learned for task A to train a related task B.
Since our attention classification is based on the gaze of the operator, gaze estimation is an ideal source task.
The training procedures for the gaze estimation and attention recognition models are described below.
The deep learning models are implemented using TensorFlow and trained on NVIDIA GeForce GTX 1060 6GB GPU.

\subsection{Datasets}
\subsubsection{ETH-XGaze}
\label{subsec:ethgaze}
ETH-XGaze~\cite{xgaze} is a large image dataset available for training gaze estimation models. 
The dataset contains over a million high-resolution images collected from $110$ participants varying in gender, age, ethnicity, etc.
It provides variations in gaze including extreme gaze angles, head poses, and differences in illumination.
The ground-truth labels for $80$ participants are available for training models, while the remaining labels (test set) have been withheld.

\subsubsection{Visual Attention Dataset}
\label{subsec:visualdata}
The \textit{Visual Attention dataset} is composed of images collected from $8$ adult volunteers ($3$ females and $5$ males, age: $18-34$ years). A Logitech C920 Pro HD webcam camera is set up in front of the table, on the available support structure, around $1.5$ meters far from the worker, as shown in Figure~\ref{fig:cell}. Each participant is asked to stand in front of the camera and to look either towards the cobot, the work table, or anywhere else, with different configurations of their head orientation and gaze direction. For each of the three conditions, $30$ pictures ($1920 \times 1080$) per person are collected and labeled accounting for a total of $720$ images.


\subsection{Gaze Estimation Model}
We use the ETH-XGaze dataset (see Section~\ref{subsec:ethgaze}) for training a gaze estimation model.
We use the VGG16~\cite{vgg16} neural network architecture pre-trained on ImageNet~\cite{imagenet}.
The VGG16 network is connected to a fully connected layer, followed by a prediction layer for estimating the gaze direction (pitch and yaw). 
We use SGD optimizer (learning rate = $0.001$), a batch size of $32$, and Mean Absolute Error as the loss function. 
The input images are scaled to the standard VGG16 input dimension of $224 \times 224$.
From the labeled dataset ($80$ participants), we reserve $8$ participants for validation and the remaining are used for training.
If the validation loss does not decrease for $5$ epochs, we reduce the learning rate by a factor of $0.1$.
To avoid over-fitting, we employ an early-stopping mechanism, i.e., we stop the training if the validation loss has not decreased in the last $7$ epochs.

\subsection{Transfer Learning Attention Recognition}
We employ a transfer learning technique for re-using the weights learned by the gaze estimation model.
That is, we copy the weights of the gaze estimation model and freeze the convolutional layers.
The weights of the frozen layers are not modified in further training steps.
We modify the prediction layer to classify the input image into three classes (cobot, table, and distracted).

The attention classes we chose are tailored to our setup.
We use the Visual Attention Dataset to map the gaze direction to the defined areas using transfer learning. 
Here we use SGD optimizer (learning rate = $0.01$) with a batch size of $15$ and Categorical Cross-Entropy as the loss function.

The input images are first cropped based on MediaPipe face detection~\cite{blazeface}.
Then, they are scaled to VGG16 input dimensions ($224 \times 224$).
During attention recognition training, we modify the brightness of the image in the range $\pm 25\%$.

We train and evaluate our model using leave-one-subject-out (LOSO) validation.
This yields a total of $8$ models (one per participant).
We calculate the Accuracy and F1-score as the measures of attention recognition performance.

\section{Evaluation}
The LOSO validation measures how well the models recognize the attention of previously unseen participants.
In addition, we explore the robustness of such a model during human-robot collaborative tasks.
To this end, we collected and annotated video data of participants working with the cobot on a collaborative assembly task.
We use this dataset to validate our gaze-based attention recognition model.

\subsection{Assembly Task Dataset}
\label{subsec:assemblydata}
Our \textit{Assembly Task Dataset} was collected in a lab setting that emulates an Industry 4.0 assembly process using the setup described in Section~\ref{sec:task}.
We collected video data from $8$ adult healthy participants ($5$ females and $3$ males, age: $18-30$ years).
Each participant is asked to work as an operator on the task for $3.5$ hours a day, for $5$ consecutive days.
This simulates the experience of an operator working a week.

We aimed to obtain a naturalistic human-robot collaboration dataset and hence, we did not guide the gaze of the participants.
Using the Logitech C920 Pro HD webcam camera, three sessions of approximately $10$ minutes each are recorded ($1280 \times 720$, $25$ fps) during the first workday (beginning, middle, and end of the workday). 
Likewise, three additional videos are acquired during the last workday of the experiment. 
With this approach, one hour of videos for each participant, for a total of $8$ hours of recording, is now available and exploited in this study to test and validate the gaze-based attention recognition model.

\subsection{Annotations}
\begin{figure*}[th]
    \includegraphics[width=0.95\textwidth]{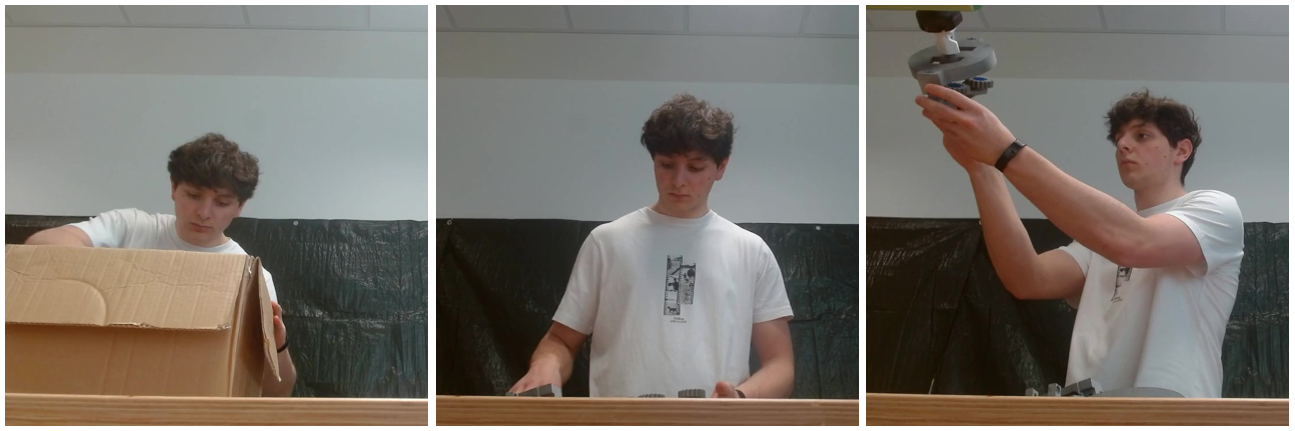}
    \caption{An image sequence visualizing the collaborative assembly task. From left to right - the operator brings the parts to the table (gathering parts), then proceeds to assemble their part of the assembly, and finally the operator and cobot collaboratively join the sub-assemblies. }
    \label{fig:assembly_sequence}
\end{figure*}

The assembly task has mainly three phases - Gathering parts, Assembly, and Collaborative joining of sub-assemblies.
Figure~\ref{fig:assembly_sequence} depicts one assembly cycle involving these three phases.
The assembly task is shared equally between the cobot and the human operator.
However, there is a disparity in speed between the cobot and the operator which results in considerable waiting time.
This leads to the operator either looking at the cobot or in a random direction (looking at the watch, window, etc.)
We annotated the videos indicating the activity the operator is doing. 
We label the waiting sequence depending on whether the operator is looking at the cobot or elsewhere (distracted).

\subsection{Test Set}
\begin{figure*}[th]
    \includegraphics[width=0.85\textwidth]{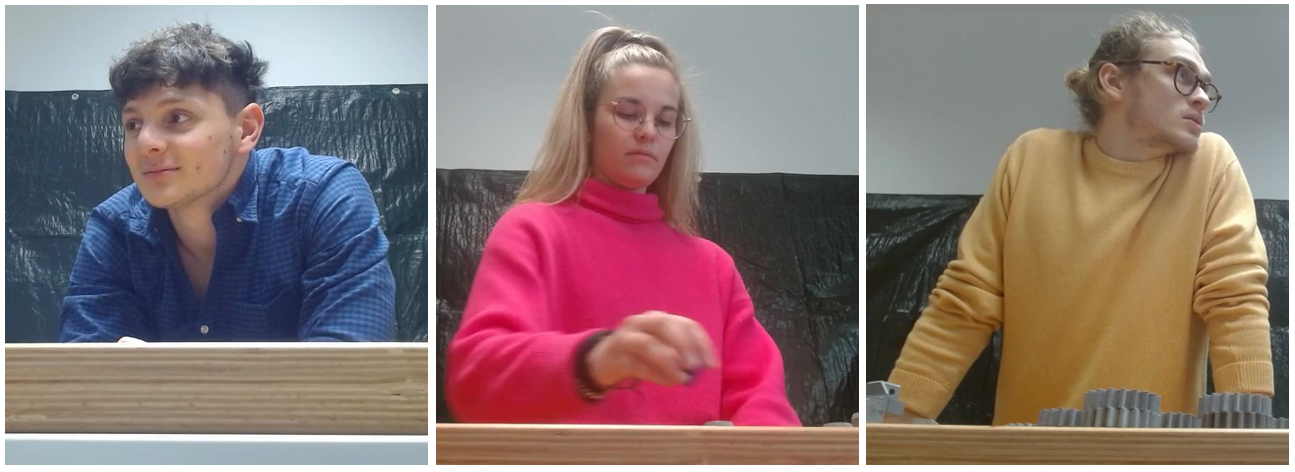}
    \caption{Some example test images from the Assembly Task Dataset. From left to right - Attention to the cobot (while waiting), Attention to the table (while assembling), and Distracted (while waiting)}
    \label{fig:assembly_dataset}
\end{figure*}

Naturally, some of the labels can be mapped to our attention classification.
During the assembly phase, the operator typically looks down at the table.
Similarly, the waiting labels can be mapped to attention on cobot and distracted classes.
Figure~\ref{fig:assembly_dataset} shows examples of images from the Assembly Task Dataset that are labeled depending on the context.

Each video segment corresponding to a label lasts a few seconds.
There is very little variance within a segment.
Hence, we choose three representative frames from a segment - the first, middle, and last frames.
First, we detect the face and crop the image. 
We reject the image if face detection failed. 
We noticed that because of the movements, some of the images were blurry.
Additionally, in a few cases, the face was occluded by some objects.
We reject the images where the eyes are not clearly visible.
This yields a test set with $833$ images of attention to cobot, $940$ images of attention on the table, and $962$ images of distraction.

\section{Results and Discussions}
In this section, we discuss the results of evaluations conducted on our attention recognition models. The results of LOSO validation are presented in Table~\ref{tab:loso}. 
Model$<i>$ refers to the model trained using the data of $i^{th}$ participant as the validation set and the remaining as the training set. 
All the models perform well in LOSO validation scoring an average of Accuracy $ = 94.3\%$ and F1-score $ = 94\%$. 
Model3 performs the best (Accuracy $ = 99\%$, F1-score $ = 99\%$). 

\begin{table*}[ht]
  \caption{Results of LOSO Evaluation on Visual Attention Dataset}
  \label{tab:loso}
  \begin{tabular}{lccccc}
    \toprule 
    Model & Recall Cobot & Recall Table & Recall Distracted & Accuracy & F1-score\\
    \midrule
     Model1 & 1.0 & 1.0 & 0.90 & 0.97 & 0.97 \\
     Model2 & 0.97 & 0.87 & 0.90 & 0.91 & 0.91 \\
     Model3 & 1.0 & 1.0 & 0.97 & 0.99 & 0.99 \\
     Model4 & 0.97 & 1.0 & 0.97 & 0.98 & 0.98 \\
     Model5 & 0.93 & 1.0 & 0.93 & 0.96 & 0.95 \\
     Model6 & 0.90 & 1.0 & 0.89 & 0.93 & 0.93 \\
     Model7 & 0.83 & 0.90 & 0.74 & 0.83 & 0.82 \\
     Model8 & 1.0 & 1.0 & 0.90 & 0.97 & 0.97 \\
    \bottomrule
     Average & & & & 0.943 & 0.94 \\
    \bottomrule
  \end{tabular}
\end{table*}

We evaluate these models in a collaborative assembly scenario using the Assembly Task Dataset. 
The results of this evaluation are presented in Table~\ref{tab:validation}.
Although the recalls are slightly lower than the Visual Attention Dataset, the models perform well and are robust in the assembly setting.
All the models have similar performance and achieve an Accuracy and F1-score of $81 - 82\%$.
\begin{table*}[ht]
  \caption{Results of Evaluation on Assembly Task Dataset}
  \label{tab:validation}
  \begin{tabular}{lccccc}
    \toprule
    Model & Recall Cobot & Recall Table & Recall Distracted & Accuracy & F1-score\\
    \midrule
     Model1 & 0.85 & 0.98 & 0.61 & 0.81 & 0.81 \\
     Model2 & 0.87 & 0.95 & 0.66 & 0.82 & 0.82 \\
     Model3 & 0.87 & 0.95 & 0.65 & 0.82 & 0.82 \\
     Model4 & 0.83 & 0.95 & 0.67 & 0.81 & 0.82 \\
     Model5 & 0.89 & 0.98 & 0.61 & 0.82 & 0.82 \\
     Model6 & 0.86 & 0.96 & 0.62 & 0.81 & 0.81 \\
     Model7 & 0.87 & 0.96 & 0.63 & 0.82 & 0.82 \\
     Model8 & 0.83 & 0.94 & 0.67 & 0.82 & 0.82 \\
    \bottomrule
     Average & & & & 0.816 & 0.818 \\
    \bottomrule
  \end{tabular}
\end{table*}

\begin{figure*}[th]
    \includegraphics[width=0.55\textwidth]{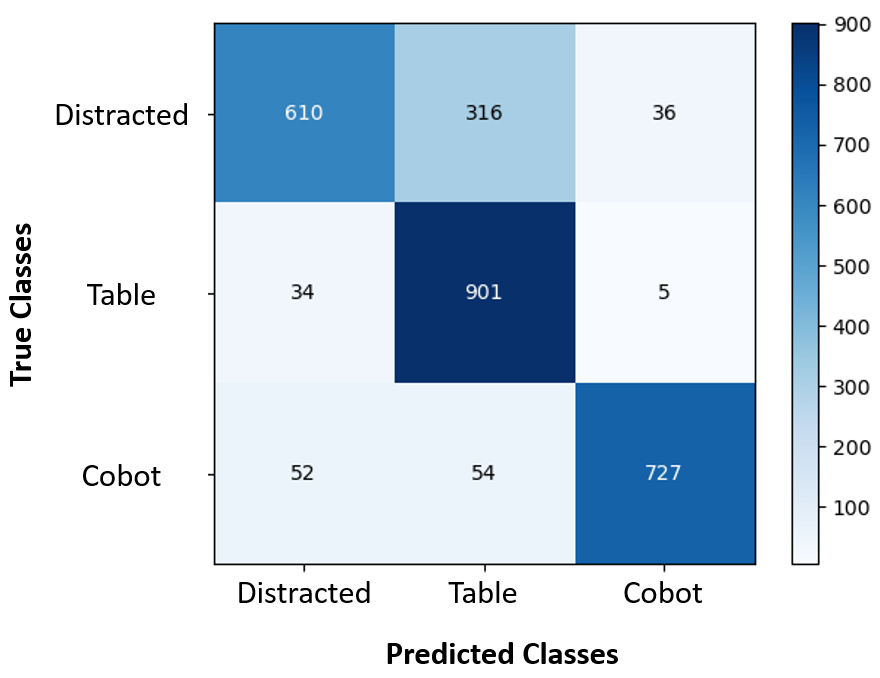}
    \caption{Confusion matrix for Model7 predictions on Assembly Task Dataset}
    \label{fig:conf_matrix}
\end{figure*}

In all the models, the Distracted class is the main contributor to the reduction in performance, followed by the Attention to Cobot class.
To gain insights into this reduction in the recall, we first investigate the confusion matrix of predictions from the models.
In Figure~\ref{fig:conf_matrix}, we visualize the confusion matrix corresponding to Model7 predictions on Assembly Task Dataset.
Interestingly, most of the misclassified Distracted images are predicted as Attention to the table.
All the models yield similar confusion matrices.

\begin{figure*}[th]
    \includegraphics[width=0.95\textwidth]{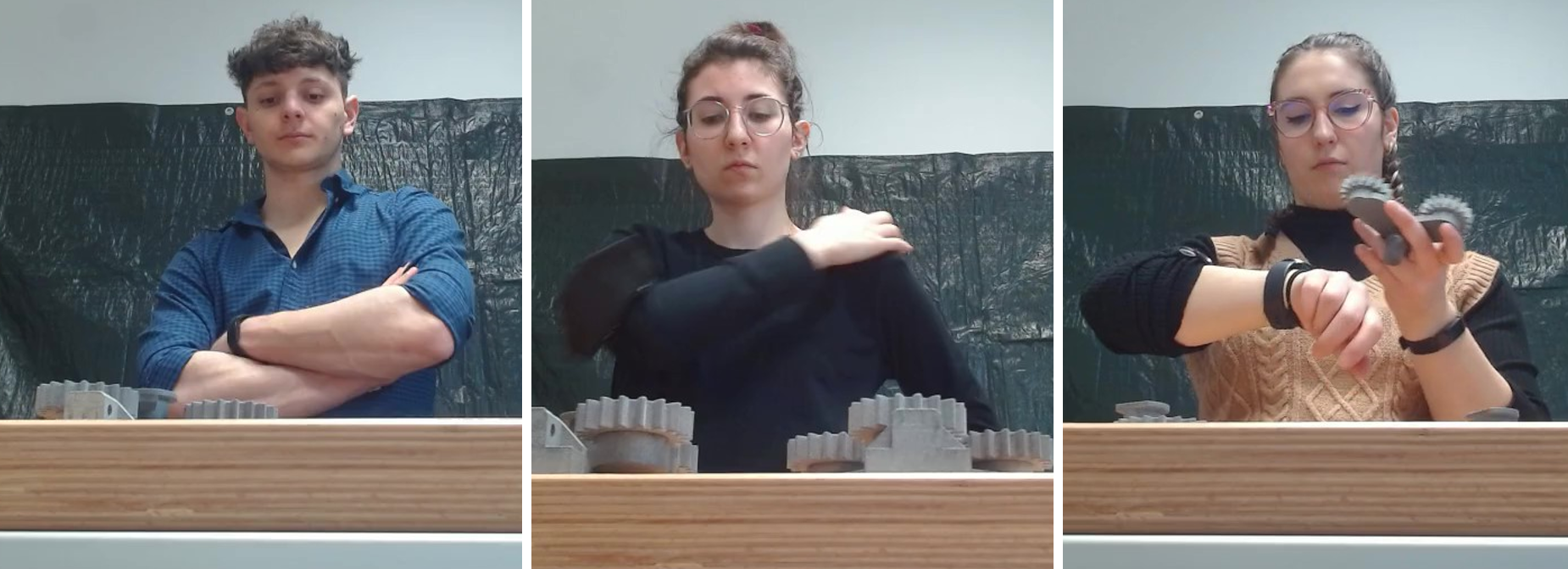}
    \caption{Few examples of Distracted images misclassified as attention to table}
    \label{fig:misclassify_examples}
\end{figure*}

We further investigate this trend by manually inspecting the misclassified images.
We observed that, in many cases, the participants were distracted by items on the table. 
For example, the participants often look at the sub-assembly on the table while waiting.
Figure~\ref{fig:misclassify_examples} shows some examples of Distracted images where the participants are looking in the direction of the table even though they are not assembling parts.
Based on our inspection, we hypothesize that the classification performance could be further improved by incorporating additional data such as the proximity of the hand to the table, and the body pose of the operator.

Through this study, we demonstrate the effectiveness of our gaze-based attention recognition model in an industrial scenario involving human-robot collaboration.
Typically studies~\cite{tayibnapis2018driver, saran2018human, huang2016anticipatory} evaluate their models in a setup where the participant is stationary and their gaze behavior is guided (specifically asked to look at certain objects/areas).
This type of evaluation is more similar to the LOSO evaluation that we performed using the Attention Detection Dataset.
However, such an evaluation lacks variances in gaze, extreme head positions, etc.
Hence, we conducted an additional evaluation of our model on the Assembly Task Dataset which is derived from human-robot collaboration sessions.
This resulted in a test set containing varying head poses, distance from the camera, etc.
We did not envision the participants getting distracted by sub-assemblies on the table.
Our observations highlight the need for evaluating attention recognition models in a non-guided setting.

\section{Conclusion}
In this work, we presented a human-robot collaborative assembly scenario that has been inspired by an Industry 4.0 use case.
As a first step towards human-centered adaptation of cobot behavior, we developed a gaze-based attention recognition model.
Based on our assembly task setup, we defined areas of interest as recognition classes - cobot, table, and anywhere else (distracted).
We collected a small dataset where the participants were instructed to look in these directions.
We used this dataset to map the gaze directions to areas of interest by performing transfer learning on a gaze estimation model.
In a leave-one-subject-out evaluation, our models performed well, achieving an average Accuracy and F1 score of approx. $94\%$.
We further evaluated our models using video snippets of participants from week-long assembly sessions using our HRC setup.
Our models again performed well, but the recall of the Distracted class was lower.
Upon manual inspection of the dataset, we found that there were many instances where the participants were distracted by sub-assemblies on the table (i.e., they were looking at the table even when they were not assembling).
Our observations highlight the need for validating attention models without any guided gaze behavior, a step that is usually missing in prior works.

Our transfer learning approach demonstrates that it is feasible to recognize human attention by fine-tuning pre-trained models, which can be deployed in realistic application scenarios where human operators work on assembly tasks with a cobot. 
The investigation of the benefit of gaze-based attention recognition in an interactive real-time setting with a cobot will be part of our future research. 
We also plan to investigate a multi-modal approach using additional non-intrusive data to improve our attention recognition. 

\begin{acks}
This work is funded by the European Union’s Horizon 2020 research and innovation programme under grant agreement No 847926 MindBot.
\end{acks}

\bibliographystyle{ACM-Reference-Format}
\bibliography{sample-base}

\end{document}